\documentstyle[aps,manuscript]{revtex}
\begin{document}

\title{Field Dependent Specific-Heat of Rare Earth Manganites}

\author{M. Roy$^{1}$, J. F. Mitchell$^{2}$, S. J. Potashnik$^{1}$, 
P. Schiffer$^{1}$\renewcommand{\thefootnote}{\alph{footnote}}
\footnote{schiffer.1@nd.edu}}

\address{$^{1}$Department of Physics, University of Notre Dame, IN 46556}
 
\address{$^{2}$Material Science Division, Argonne National Laboratory,
 Argonne, IL 60439}

\maketitle
\begin{abstract}

{\hspace{0.5cm} The low temperature specific heat C(H) of several rare-earth manganites (La$_{0.7}$Sr$_{0.3}$MnO$_{3}$, Nd$_{0.5}$Sr$_{0.5}$MnO$_{3}$, 
Pr$_{0.5}$Sr$_{0.5}$MnO$_{3}$, La$_{0.67}$Ca$_{0.33}$MnO$_{3}$, 
La$_{0.5}$Ca$_{0.5}$MnO$_{3}$, La$_{0.45}$Ca$_{0.55}$MnO$_{3}$ and
La$_{0.33}$Ca$_{0.67}$MnO$_{3}$) was measured as a function of magnetic
field. We observed behaviour consistent with thermodynamic
expectations, i.e., C(H) decreases with field for ferromagnetic metallic 
compounds by an amount which is in quantitative agreement with
spin wave theory. We also find that C(H) increases with field in  
most compounds with a charge-ordered antiferromagnetic ground state. In 
compounds which show evidence of a coexistence of ferromagnetic 
metallic and antiferromagnetic charge-ordered states, C(H) 
displays some unusual non-equilibrium effects presumably associated 
with the phase-separation of the two states. We also observe a large anomalous
low temperature specific heat at the doping induced metal-insulator 
transition (at x $=$ 0.50) in La$_{1-x}$Ca$_{x}$MnO$_{3}$.}    

\end {abstract}

\newpage
\newpage

The rare-earth perovskite manganites, R$_{1-x}$A$_{x}$MnO$_{3}$ (R and A
are trivalent and divalent ions respectively) exhibit some intriguing 
features originating from the strong interplay between the electronic,
magnetic and structural degrees of freedom. Recent studies have 
demonstrated that the \cite{apr:97} ground state of these
materials can be changed by varying either the band-filling 
\cite{tomi:95,woll:55,ps:95} or the band-width 
\cite{hwang:95,font:96,maig:96}, and that the behaviour is extremely 
sensitive to the application of an external magnetic field leading to the
so-called $``$colossal magnetoresistance$"$.  
   
The stoichiometric manganites (RMnO$_{3}$) are antiferromagnetic insulators
\cite{mura:98,solo:96}, but the ground state of these materials can be 
tuned to a metallic ferromagnetic state by increasing the band filling \cite{ps:95}. At larger values of x, the ferromagnetic conducting state 
becomes unstable to a charge-ordered \cite {apr:96} antiferromagnetic 
ground state with large resistivity. For some values of x, 
the charge-ordered state can be dissociated by an external magnetic 
field, leading to a large magnetoresistance of many orders of magnitude. 
It has been recently demonstrated that there is a coexistence of 
phase-separated ferromagnetic metallic (FMM) and antiferromagnetic 
charge-ordered (ACO) states at both values of x near the cross-over 
between the two ground states, and also in the lower doping regime \cite{chen:96,rada:97,papa:97,allo:97,Henn:98,calv:98,allo:98,papa1:97,yuno:98,more:99,roy:98}.
 
Measurements of specific heat provide insight into the nature of the 
excitations in these various phases and the phenomena responsible 
for the metal-insulator transitions. There have been a number of studies
of the specific heat of the manganites,
\cite{hami:96,wood:97,park:97,smol:98,lees:99,gord:99} but these studies 
primarily concentrated
on the temperature dependence of the specific heat, and the contributions of
different excitations to the specific heat were derived from analytical 
fits to the data. Since the field-dependent properties of these materials
are responsible for much of the recent interest, we have performed 
a study of the field dependence of the low temperature C(H) on a 
range of manganite samples to elucidate how the thermodynamic 
properties of rare-earth manganites evolve as a function of magnetic field. 
We observe that the magnetic specific heat dominates the behaviour
of C(H) and that the behaviour is typically consistent with 
simple thermodynamic expectations.

Specific heat was measured on both polycrystalline (pc) and single crystal (sc)
manganite samples. We studied two samples, La$_{0.67}$Ca$_{0.33}$MnO$_{3}$ 
(pc) and La$_{0.7}$Sr$_{0.3}$MnO$_{3}$ (sc) with FMM ground states, three
samples, Nd$_{0.5}$Sr$_{0.5}$MnO$_{3}$ (sc), Pr$_{0.5}$Sr$_{0.5}$MnO$_{3}$
(sc) and La$_{0.33}$Ca$_{0.67}$MnO$_{3}$ (pc) with ACO ground states and two
samples, La$_{0.5}$Ca$_{0.5}$MnO$_{3}$ (pc) and 
La$_{0.45}$Ca$_{0.55}$MnO$_{3}$ (pc) which display strong evidence of 
a coexistence of FMM and ACO states. The single crystal samples were grown by
floating-zone method, and the polycrystalline samples were synthesized by the 
solid-state reaction method. All our samples were single phase as adjudged by 
x-ray diffraction. Specific heat was measured by semi-adiabatic 
heat pulse method. All the samples were zero-field cooled to the 
prescribed temperature and the specific heat was measured as a 
function of field at a constant temperature in discrete steps of 
0.25 T, as the field was swept from 0 $\rightarrow$ 9 T, 
9 T $\rightarrow$ 0 $\rightarrow$ -9 T and -9 T $\rightarrow$ 0 
$\rightarrow$ 9 T. Each step in field took a total of $\sim$
10-30 minutes, depending on the thermal mass of the samples,
resulting in some non-equilibrium effects as discussed below. 
Moreover, the measured values of C(T) are in good agreement with
those of previous studies \cite{hami:96,wood:97,park:97,smol:98,lees:99,gord:99}.

From simple thermodynamics, one expects the field dependence of the
magnetic contribution to the specific heat to be given by:

\begin{equation}
{\left(\frac{\partial{C}}{\partial{H}}\right)_{T} 
\equiv T\left(\frac{\partial^{2}{M}}{\partial{T^{2}}}\right)_{H}                               }
\label{e3}
\end{equation} 
For a ferromagnet in a positive field 
$\frac{\partial^{2}M}{\partial T^{2}}$ $<$ 0, thus one expects 
$\frac{\partial C}{\partial H}$ $<$ 0, i.e., C(H) decreases in a field.
This can also be interpreted as the suppression of spin waves
with increasing field. By contrast, for an antiferromagnet
in a positive field, $\frac{\partial^2{M}}{\partial{T^2}}$ $>$ 0
$\Longrightarrow$ $\frac{\partial{C}}{\partial{H}}$ $>$ 0, i.e., 
C(H) increases with increasing field, which can alternately be viewed
as the softening of the AFM order parameter in a field.

Figure \ref{f1} shows C(H) of
La$_{0.7}$Sr$_{0.3}$MnO$_{3}$ (sc) and La$_{0.67}$Ca$_{0.33}$MnO$_{3}$ (pc)
at T = 5.45 K. The specific heat of both samples decreases 
monotonically with increasing field, dropping by
$\sim$ 10 $\%$ as the field is changed from 0 to 9T, although C(H) of
La$_{0.67}$Ca$_{0.33}$MnO$_{3}$ is $\sim$  18$\%$ higher, attributable to the 
larger lattice contribution due to the smaller size of Ca ion. The
reduction in C(H) with increasing field can be explained as the
suppression of thermal excitations of the spin waves as described above. 
Similar observations were also drawn from the recent temperature dependent 
specific-heat experiments \cite{okud:98} on La$_{1-x}$Sr$_{x}$MnO$_{3}$.

The magnon contributions to the specific heat of a ferromagnet 
in an external field neglecting any demagnetization effects, and 
assuming no spin gap at H = 0, so that the dispersion relation at an
external field H is $\omega = g\mu_{B}H + Dk^{2}$, can be written as \cite{kitt:64}, 

\begin{equation}
{C_{magnon}(H) = \frac{k_{B}^{5/2}T^{3/2}}{4\pi^{2}D^{3/2}}
\int\limits_{g\mu_{B}H/k_{B}T}^{\infty}
\frac{x^2 e^{x}}{{({e^{x}-1})}^2}{\sqrt{x-\frac{g\mu_{B}H}{k_{B}T}}}dx} 
\label{e2}
\end{equation}

where $x = (g\mu_{B}H+Dk^{2})/k_{B}T$, and D = 2JS$a^{2}$ is the spin 
stiffness constant. We can thus calculate the spin-wave contributions to the 
specific heat of these two samples from calculations based on Eq. 
\ref{e2}, and using stiffness constant \cite {endo:97,lynn:96} and 
unit cell length \cite{mahe:96} from previous neutron scattering 
and x-ray scattering data respectively. As shown by the solid lines 
in figure \ref{f1}, the spin wave calculation fits the C(H) data 
extremely well with only a single free parameter -- a constant offset 
which accounts for the non-magnetic contributions to the heat capacity.

Figure \ref{f2} and \ref{f3} show the low temperature C(H) of single crystals 
of Pr$_{0.5}$Sr$_{0.5}$MnO$_{3}$ (PSMO) and Nd$_{0.5}$Sr$_{0.5}$MnO$_{3}$
(NSMO). Recent neutron and x-ray scattering experiments \cite{kawa:97} have 
suggested that at low temperatures while the orbital ordering of d${_{x^{2}}^{2}}$-d${_{y^{2}}^{2}}$ leads to A-type
AFM state in PSMO with no clear evidence for a charge ordering, 
d${_{z^{2}}}$ orbitals order in NSMO resulting in a CE-type 
($\pi$, 0, $\pi$) type charge ordered AFM state. However, PSMO 
undergoes a sharp rise in resistivity with an accompanying 
drop in magnetization, which is usually associated
with charge-ordering. At moderate fields the low temperature C(H) of
both PSMO and NSMO increases monotonically with increasing field in 
contrast to the FM samples, but consistent with the expectations for
the AFM materials. The low temperature C(H) of PSMO (figure \ref{f2}) 
increases monotonically with field at low fields due to the softening 
of the stiffness constant in the AFM state. 
However, in the region (4 T $\lesssim$ H $\lesssim$ 8 T), C(H) has 
a reduced slope, suggesting that the AFM $\rightarrow$ FM transition 
which occurs at $\sim$ 5 T \cite{tomi:95} is not completed until H 
$\sim$ 8 T. At higher fields (H $\gtrsim$ 8 T), in contrast to NSMO 
(see below) we believe that the change in C(H) with field is 
electronic in nature, and the steep rise in C(H) with increasing field, 
which coincides with the sharp drop in $\rho(H)$ is due to the enhancement 
of the free carriers as the FM state is percolated throughout the 
sample. This sample also shows some non-equilibrium effects, e.g., C(H) 
increases by 8 $\%$ in $\sim$ 10 mins immediately after the field is raised
to 9 T, even when the external parameters, such as temperature and the 
field remained unchanged. This kind of non-equilibrium behaviour
is much more pronounced in the sample which shows strong evidence of 
a coexistence of FM and charge-ordered state, as discussed below. 
The gradual relaxation of the FM state in this sample is also
possibly responsible for the smaller C(H) (by $\sim$ 10$\%$) at 
low fields during subsequent field sweeps.

The low temperature specific heat of NSMO is an 
order of magnitude higher than the rest of the measured manganites, this
is possibly associated with Schottky-like anomaly connected with 
the large magnetic moment (J $=$ 9/2) of the Nd$^{3+}$ ion 
\cite{gord:99,coey:95}. The low temperature C(H) of NSMO increases 
monotonically with field at H $\lesssim$ 7 T, demonstrating a stable 
\cite{kawa:97,mahe:99} AFM state. At H $\gtrsim$ 7 T, the slope 
of C(H) changes, and C(H) displays a sharp downward turn 
(see figure \ref{f3}), suggesting 
the suppression of thermal excitation of the spin-waves as the sample 
approaches an AFM $\rightarrow$ FM transition. During 
subsequent field sweeps M(H) shows no hysteresis, although $\rho(H)$ 
remains order of magnitude smaller, indicating that the carriers 
remain delocalized even when the field is removed. Since C(H) shows no 
hysteresis despite the increase in the population of free carriers 
\cite{roy:98}, this further suggests that the magnetic-contribution 
dominates the low temperature heat capacity. 

Figure \ref{f4} shows the low temperature C(H) of
La$_{0.33}$Ca$_{0.67}$MnO$_{3}$. This sample exhibits
all the features of an ACO ground state, with 
magnetization reaching only $\sim$ 2 $\%$ of the total
saturation magnetization even at H $=$ 7 T. However, the
behaviour of C(H) is quite different from the other AFM CO 
samples we studied. The low temperature C(H) of this sample 
exhibits features which are rather inconsistent with the 
thermodynamical expectations for an AFM state, i.e, C(H) decreases
monotonically with increasing field, although the slope of
C(H) is not as sharp as that of the FMM samples. However, when 
the field is increased from 0 to 9 T, C(H) drops by only 
3.5 mJ. The absolute magnitude of this change is 
at least a factor of 3 smaller than that of the other measured 
AFM samples, but quantitatively and qualitatively similar to the
equilibrium C(H) of the phase-separated samples, as discussed 
below. This uncharacteristic behaviour of C(H) in the AFM state
can not be attributed to the magnetic excitations, and is
perhaps associated with charge/orbital ordering 
of the La$_{1-x}$Ca$_{x}$MnO$_{3}$ compounds.

Figure \ref{f5} and \ref{f6} show low temperature C(H) of  
polycrystalline sample of La$_{0.50}$Ca$_{0.5}$MnO$_{3}$ 
(Mn$^{4+}\%$ = 53.8) and  La$_{0.45}$Ca$_{0.55}$MnO$_{3}$  
(Mn$^{4+}\%$ = 58.2) with well characterized Mn$^{4+}$ content. 
These samples show strong evidence of phase separation into FMM 
and insulating ACO states \cite{chen:96,rada:97,papa:97,allo:97,Henn:98,calv:98,allo:98,papa1:97,yuno:98,more:99,roy:98}. The charge-ordered state of the sample with 
x $=$ 0.50 can be partially dissociated by a moderate external magnetic 
field, greatly reducing the resistivity, but the sample with x $=$ 0.55 
remains largely charge-ordered showing very little magnetoresistance
even at H $=$ 9 T. 

We find that the behaviour of C(H) in the phase-separated samples
is more complex than the others we have studied, displaying
significant non-equilibrium effects. The low temperature zero-field 
cooled C(H) of La$_{0.5}$Ca$_{0.5}$MnO$_{3}$ decreases with increasing 
field, when the field is swept for the first time from H $=$ 0 $\rightarrow$
9 T. After reaching 9 T, however, C(H) increases by $\sim$ 19$\%$ in
$\sim$ 10 mins even when field is kept constant at 9 T. On decreasing the 
field, C(H) not only remains 17$\%$ higher than the initial sweep but is 
almost constant for H $\gtrsim$ 3 T. However, at lower fields, C(H) drops 
sharply at H $\sim$ 3 T with a minimum at H $\sim$ 1 T, and then
recovers to near its maximum value at H $\sim$ -1 T. Although the 
low temperature C(H) changes by only $\lesssim$ 3$\%$ on increasing 
the field in the reverse direction for H $\gtrsim$ -6 T, C(H) drops 
sharply at H $\sim$ -6 T, and similar features to those of the 
initial sweep are observed for H $\lesssim$ -6 T and also during
subsequent field sweep from H = -9 T $\rightarrow$ 9 T. To characterize
equilibrium behaviour of C(H), we also performed the same measurements
in an equilibrium mode, i.e., data were taken after
waiting for 1 hour at every field, ensuring that the system 
had reached equilibrium at that field by monitoring C as a 
function of time at every field. We observe that the data
taken in this equilibrium mode displays only $\sim$ 2$\%$ rise in C(H) 
at H $\lesssim$ 3 T, but at higher fields remains almost flat up to 9 T.

The sharp rise in C after the field is first swept to 9 T suggests 
that the FM regimes are growing with time at the expense of the ACO 
phase. This increase in C could then be attributable to an increased
population of free carriers or as the enhancement of long wavelength
low energy spin-wave excitations with the percolation of the small
FM regimes. The minima we observe in C(H) at $\sim$ 1 T when 
sweeping the field could similarly be explained by the gradual 
domain formation or a time-dependent increase in the fraction of 
the sample which is phase separated into the ACO phase. That C(H) 
measured in equilibrium changes very little with field is consistent
with behaviour of the La$_{0.33}$Ca$_{0.67}$MnO$_{3}$ sample. This
suggests that the magnetic heat capacity of the CO state in
La$_{1-x}$Ca$_{x}$MnO$_{3}$ is much smaller than that of the other
ACO samples studied. This is perhaps due to the microscopic 
phase-separation which inhibits the formation of long wavelength 
spin modes in La$_{1-x}$Ca$_{x}$MnO$_{3}$ systems \cite{mori:98}. 
 
The charge-lattice of La$_{0.45}$Ca$_{0.55}$MnO$_{3}$ remains 
primarily intact even at H $\sim$ 9 T. Although magnetization 
reveals the presence of small clusters of ferromagnetism, which undergo 
a first order transition to charge-ordering at a lower temperature 
than the bulk of the sample, a FM state is not established throughout
the sample even at H $=$ 9 T \cite{roy:98}. On increasing the field, 
the low temperature C(H) increases by 7$\%$ but C(H) displays a 
small downward turn at H $\sim$ 4 T before finally dropping sharply 
at H $\sim$ 6 T. After reaching 9 T, C(H) rises by 26$\%$ in
$\sim$ 10 minutes even when the external parameters remained unchanged. 
Similar features are also observed during subsequent field sweeps 
though C(H) shows no non-equilibrium effect. The 
non-equilibrium effects appear to be intrinsic 
to the samples which display a coexistence of small clusters of
ferromagnetism in the primarily charge-ordered regime, although 
the extent of this effect decreases for samples far from x $=$ 0.50.

One other anomalous feature of the heat capacity of the La$_{0.5}$Ca$_{0.5}$
MnO$_{3}$ sample is that the magnitude of C(H=0) is much larger than that of 
La$_{0.67}$Ca$_{0.33}$MnO$_{3}$ or La$_{0.33}$Ca$_{0.67}$MnO$_{3}$.  In 
fact the enhancement of heat capacity at x $=$ 0.50 has been
previously noted, and was suggested to be due to an extra contribution
to C by orbital excitations of the d$_{z^2}$ orbital ordering 
\cite{smol:98}. Since we have available samples of 
La$_{1-x}$Ca$_{x}$MnO$_{3}$ for a wide range of x \cite{roy:98}, we also 
studied this phenomenon briefly with the results plotted in figure 
\ref{f7}. We see that there is a sharp rise in C(x) for x $=$ 0.5 with 
some scatter around the maximum. This scatter is greatly reduced, 
however when the data are plotted as a function of the actual Mn$^{4+}$ 
content of the samples (as determined by redox titration).  This 
maximum appears to be associated with the doping-induced 
metal-insulator transition at x $\sim$ 0.50, and since it is not 
reflected in the field-dependence of the heat capacity, 
and since the lattice properties of these materials are quite similar. 
We hypothesize that the enhancement of C(x) near x $=$ 0.50 
\cite{toku:93,kuma:93,toku1:93,fuji:92,moriy:95} is 
associated with an enhancement of the electron mass near the MIT, 
but there is no detailed theory to support this possibility.

In conclusion, we have measured low temperature field dependent 
specific heat of a range of rare earth manganites. Our data 
suggest that the behaviour of C(H) is dominated
by the magnetic contributions to the specific heat regardless of the
nature of the magnetic or electronic state  of these materials.
Moreover, we observed that the slope of C(H) with respect to
the field for both ferromagnetic and antiferromagnetic samples
is typically consistent with the thermodynamic expectations. We also 
observed that C(H) of compounds which exhibit evidence of a
coexistence of ferromagnetic metallic and antiferromagnetic 
charge-ordered states show some unusual non-equilibrium effects
presumably associated with the phase-separation of the two states.

We are grateful to Dr. A. P. Ramirez for many enlightening discussions. 
This research has been supported by NSF grant DMR 97-01548,
the Alfred P. Sloan Foundation and the Dept. of Energy, Basic Energy
Sciences-Materials Sciences under contract $\#$W-31-109-ENG-38.

\begin{figure}

\caption {\label {f1} The specific heat of La$_{0.7}$Sr$_{0.3}$MnO$_{3}$
and La$_{0.67}$Ca$_{0.33}$MnO$_{3}$ as a function of field at 
T = 5.5 K, when the field was swept from 0 $\rightarrow$ 9 T (open circles),
9 T $\rightarrow$ -9 T (solid lines) and -9 T 
$\rightarrow$ 9 T (dashed line). The thick solid line is based on calculation
using Eq. \ref{e2}, as explained in the text. A constant offset is
added to the calculated value to account for the non-magnetic contributions
to C(H).}

\end {figure}

\begin{figure}

\caption {\label {f2} The specific heat of Pr$_{0.5}$Sr$_{0.5}$MnO$_{3}$
at T= 5.5 K as a function of field.  All the measurements were done 
when the field was swept from 0 $\rightarrow$ 9 T (open circles), 
9 T $\rightarrow$ -9 T (solid lines) and -9 T $\rightarrow$ 9 T
(dashed line). The top and the bottom panel of the
inset show resistivity and magnetization at T = 5 K as a function 
of field.}

\end{figure}

\begin{figure}

\caption {\label {f3} The specific heat of Nd$_{0.5}$Sr$_{0.5}$MnO$_{3}$
at T = 5.5 K as a function of field. All the measurements were done when the field was swept from 0 $\rightarrow$ 9 T (open circles), 9 T 
$\rightarrow$ -9 T (solid lines) and -9 T $\rightarrow$ 9 T 
(dashed line). The top and the bottom panels of the 
inset show resistivity and magnetization at T = 5 K as a function of 
field. }

\end{figure}

\begin{figure}
  
\caption {\label {f4} The specific heat of La$_{0.33}$Ca$_{0.67}$MnO$_{3}$ 
as a function of field at T = 5.5 K, when the field was swept 
from 0 $\rightarrow$ 9 T (open circles), 9 T $\rightarrow$ -9 T 
(solid lines) and -9 T $\rightarrow$ 9 T (dashed line).}

\end {figure}

\begin{figure}

\caption {\label {f5} The specific heat of La$_{0.50}$Ca$_{0.50}$MnO$_{3}$
as a function of field at T = 5.5 K, when the sample was measured in
continuous mode, when the field was swept from 0 
$\rightarrow$ 9 T (open circles), 9 T $\rightarrow$ -9 T 
(solid lines) and -9 T $\rightarrow$ 9 T 
(dashed line), and at equilibrium (increasing the field (solid down triangles)
and decreasing the field (solid up triangles)) as discussed in the 
text. The left and right panels of the inset show resistivity 
($M\Omega$ cm) and magnetization at T = 50 K and T = 5 K respectively as 
a function of field.}

\end {figure}

\begin{figure}
  
\caption {\label {f6} The specific heat of La$_{0.45}$Ca$_{0.55}$MnO$_{3}$ 
as a function of field at T = 5.5 K, when the field was swept 
from 0 $\rightarrow$ 9 T (open circles), 9 T $\rightarrow$ -9 T 
(solid lines) and -9 T $\rightarrow$ 9 T (dashed line). The inset illustrates
the field dependent magnetization at T $=$ 5 K. }

\end {figure}

\begin{figure}
  
\caption {\label {f7} The specific heat at H $=$ 0 of 
La$_{1-x}$Ca$_{x}$MnO$_{3}$ as a function of Ca doping and measured
Mn$^{4+}$ content at T $=$ 5.5 K. The large stars illustrate the data from
previous measurements from [22,25].} 

\end {figure}

\end{document}